\documentstyle[11pt,aaspp4,epsf,flushrt]{article}

\makeatletter
\def\thebibliography#1{\section*{REFERENCES}
 \addcontentsline{toc}{section}{REFERENCES}
 \list{}{\labelwidth\z@
         \leftmargin 1.5em
	 \itemsep \z@
	 \itemindent-\leftmargin}
 \small\raggedright
 \parindent\z@
 \parskip\z@ plus .1pt\relax
 \def\newblock{\hskip .11em plus .33em minus .07em}
 \sloppy\clubpenalty4000\widowpenalty4000
 \sfcode`\.=1000\relax
}

\def\@biblabel#1{}
\def\@bcite#1#2{(#1\if@tempswa , #2\fi)}
\def\@pcite#1#2{#1\if@tempswa , #2\fi}
\def\@citefmta#1#2{#1 (#2)}
\def\@citefmtb#1#2{#1 #2}
\let\citefmt=\@citefmta

\def\@citex[#1]#2{\if@filesw\immediate\write\@auxout{\string\citation{#2}}\fi
  \def\@citea{}\@cite{\@for\@citeb:=#2\do
    {\@citea\def\@citea{;\penalty\@m\ }\@ifundefined
    {b@\@citeb}{{\bf ?}\@warning
{Citation `\@citeb' on page \thepage \space undefined}}%
{\csname b@\@citeb\endcsname}}}{#1}}

\def\cite{\@ifnextchar [{\let\citefmt=\@citefmtb
                          \let\@cite=\@bcite\@tempswatrue \@citex}
                        {\let\citefmt=\@citefmtb
                          \let\@cite=\@bcite\@tempswafalse \@citex[]}}
\def\pcite{\@ifnextchar [{\let\citefmt=\@citefmtb
                          \let\@cite=\@pcite\@tempswatrue\@citex}
                        {\let\citefmt=\@citefmtb
                          \let\@cite=\@pcite\@tempswafalse\@citex[]}}
\def\scite{\@ifnextchar [{\let\citefmt=\@citefmta
                          \let\@cite=\@pcite\@tempswatrue\@citex}
                        {\let\citefmt=\@citefmta
                          \let\@cite=\@pcite\@tempswafalse\@citex[]}}
\makeatother

\def\d{{\rm d}}
\def\e{{\rm e}}

\def\M{{\rm M}}

\def\Z{{\rm Z}}

\def\bx{{\mathbf{x}}}

\def\deg{^{\rm o}}
\def\Mpc{\ifmmode{{\rm Mpc}}\else{Mpc}\fi}
\def\hMpc{\ifmmode{h^{-1}{\rm Mpc}}\else{$h^{-1}$Mpc}\fi}
\def\kms{\ifmmode{{\rm km}/{\rm s}}\else{${\rm km}/{\rm s}$}\fi}

\def\realR{{{\rm I\kern-0.16em{}R}}}
\def\Vol{{\rm Vol}}
\def\Pois{{\rm P}}

\def\ViendalMare{Roberto Trasarti--Battistoni}
\def\ii{{\'\i}}

\begin{document}

\title{A global descriptor of spatial pattern interaction in the
galaxy distribution }

\author{
Martin Kerscher\altaffilmark{1,7}, 
Mar\ii a Jes\'us Pons--Border\ii a\altaffilmark{2},
Jens Schmalzing\altaffilmark{1,3}, 
\ViendalMare\altaffilmark{1,4},
Thomas Buchert\altaffilmark{1}, 
Vicent J. Mart\ii nez\altaffilmark{5},
and Riccardo Valdarnini\altaffilmark{6} }
\altaffiltext{1}{Ludwig--Maximilians--Universit\"at, Theresienstra{\ss}e 37,
80333 M\"unchen, Germany}
\altaffiltext{2}{Departamento de F\'{\i}sica Te\'orica, Universidad
Aut\'onoma de Madrid, 28049 Cantoblanco, Madrid, Spain }
\altaffiltext{3}{Max--Planck--Institut f\"ur Astrophysik,
Karl--Schwarzschild--Stra{\ss}e 1, 85740 Garching, Germany}
\altaffiltext{4}{Sezione di Astrofisica \& Cosmologia, Dipartimento di
Fisica, Univerist\`a di Milano, via Celoria 16, 20133 Milano, Italy}
\altaffiltext{5}{Departament d'Astronomia i Astrof\'{\i}sica, Universitat de
Val\`encia, 46100 Burjassot, Val\`encia, Spain}
\altaffiltext{6}{SISSA, Via Beirut 2-4, 34014 Trieste, Italy}
\altaffiltext{7}{email kerscher@stat.physik.uni-muenchen.de}

\begin{abstract}
We present  the function $J$ as a   morphological descriptor for point
patterns formed by the distribution of galaxies in the Universe.  This
function  was recently introduced in  the field of spatial statistics,
and is  based  on  the  nearest  neighbor  distribution and   the void
probability  function.  The  $J$   descriptor  allows to   distinguish
clustered (i.e.\ correlated) from ``regular'' (i.e.\ anti--correlated)
point distributions.  We  outline  the theoretical foundations  of the
method, perform tests with a Mat\'ern  cluster process as an idealised
model of  galaxy clustering, and apply the  descriptor to galaxies and
loose    groups in  the    Perseus--Pisces Survey.   A comparison with
mock--samples extracted from a mixed dark matter simulation shows that
the $J$  descriptor can be profitably  used to constrain (in this case
reject) viable models of cosmic structure formation.
\end{abstract}


\section{Introduction}

Three--dimensional  patterns formed by   the spatial  distribution  of
galaxies in the Universe have already been described and quantified by
various    methods:      correlation     functions,  counts--in--cells
{}\cite{peebles:principles},        the      void probability function
{}\cite{white:hierarchy},    the  genus  {}\cite{melott:review},   the
multifractal spectrum  {}\cite{martinez:clustering},     skewness  and
kurtosis  {}\cite{gaztanaga:bias},      and      Minkowski functionals
({}\pcite{mecke:robust},  {}\pcite{schmalzing:beyond}).  Some of these
descriptors are complementary and suggest a physical interpretation of
cosmic patterns   by  emphasising different  spatial  features  of the
galaxy distribution.

The treatment of the galaxy distribution as a realization of a spatial
point process promises   useful insights  through the  application  of
methods from   the  field  of   spatial statistics.    The forthcoming
three--dimensional galaxy   catalogues with more  than half  a million
redshifts additionally motivate  the  development of new   statistical
techniques.

In this article  we want to reinforce a  morphological measure for the
study of the distribution of galaxies, the $J(r)$--function, which has
recently  been  introduced into the    field of spatial statistics  by
{}\scite{vanlieshout:j}   and is  related   to   the nearest--neighbor
distribution  $G(r)$  and the  spherical  contact distribution $F(r)$.
Indeed,  the   $J(r)$--function  is equal   to   the first conditional
correlation       function              (\pcite{stratonovich:topicsI},
{}\pcite{white:hierarchy}), and was  used  by {}\scite{sharp:holes} to
test a  hierarchical ansatz for $n$--point  correlation functions.  We
will  focus on different features of  the $J(r)$--function showing its
discriminative power as a measure of the strength of clustering.

Our  article is organised  as  follows: In Sect.~\ref{sect:nearest} we
present the  distribution functions $F$ and  $G$ and show  how the $J$
function is constructed. A Mat\'ern cluster process is considered as a
simple example of a clustering point distribution.
In  Sect.~\ref{sect:galaxies} we study the  clustering properties of a
galaxy   sample and  of   galaxies   in   groups extracted  from   the
Perseus--Pisces redshift survey (PPS).
We   compare   the observed   galaxy   distribution with  mock samples
extracted  from  a    Mixed  Dark     Matter  (MDM)  simulation     in
Sect.~\ref{sect:nbody}.
We summarise and conclude in Sect.~\ref{sect:outlook}.

\section{The $J$ function}
\label{sect:nearest}

In the theory of spatial point processes the distribution of a point's
distance to its nearest neighbor is a common tool for the analysis of
point patterns {}\cite{stoyan:stochgeom}.   We  consider  the redshift
space coordinates $\{\bx_{i}\}_{i=1}^{N}\in\realR^{3}$ of $N$ galaxies
inside  a region $D\subseteq\realR^3$  as  a realization  of the point
process  describing  the  spatial  distribution   of galaxies  in  the
Universe.

The nearest neighbor distribution $G(r)$ is the distribution function
of the distance $r$  of a point of   the process to the  nearest other
point of the process.
Similarly,  the  spherical  contact    distribution $F(r)$     is  the
distribution function of  the distance  $r$  of an  arbitrary point in
$\realR^{3}$ to the nearest point of the process.
$F(r)$  is equal to  the  volume fraction  occupied by the  set of all
points in $D$ which  are closer than  $r$ to  a point of  the process.
Hence, $F(r)$ coincides with the volume density of the first Minkowski
functional   ({}\pcite{mecke:robust} and {}\pcite{kerscher:abell}) and
is related     to  the void   probability function     $P_{0}(r)$  via
$F(r)=1-P_{0}(r)$.

For a homogeneous Poisson process we have
\begin{equation}
\label{eq:vpf}
F_\Pois(r)=1-\exp\left(-\frac{4\pi}{3}{r^3}n\right)=G_\Pois(r) ,
\end{equation}
where $n$ is the number density.  
Boundary--corrected   estimators  for   both  the    nearest  neighbor
distribution and  the   spherical contact  distribution  used  in  our
studies are   provided   by    minus  (reduced   sample)    estimators
({}\pcite{stoyan:stochgeom},          also       detailed           in
{}\pcite{kerscher:fluctuations}).

In a recent paper, {}\scite{vanlieshout:j} have suggested to use the
quotient
\begin{equation}
J(r) = \frac{1-G(r)}{1-F(r)}
\end{equation}
for characterising a point process; in that  way the surroundings of a
point belonging to the process and  the neighborhood of a random point
are compared.  They consider several point  process models and provide
limits     and       exact   results   on         $J(r)$   (see   also
Section~\ref{sect:matern}). 

If  the process under consideration  is  clustered, an arbitrary point
usually lies farther away from a point of the process than in the case
of    a  Poisson  process.    Hence,   clustering   is   indicated  by
$F(r)<F_\Pois(r)$.  Consistently, $G(r)>G_\Pois(r)$,  since  clustered
points tend  to lie closer to their  nearest  neighbors than randomly
distributed points.  So, for a clustered point distribution, $J(r)<1$.

In case of  anti--correlated, ``regular'' structures  the situation is
the opposite: on average a point of a  regular process is farther away
from the nearest other point of the process, so $G(r)<G_\Pois(r)$, and
a  random point is  closer to  a  point of  the process,  resulting in
$F(r)>F_\Pois(r)$.    Therefore, regular structures   are indicated by
$J(r)>1$.

For a homogeneous Poisson process we obtain $J_\Pois(r)=1$, separating
regular from clustering structures.

\subsection{The Mat\'ern cluster process}
\label{sect:matern}

\begin{figure}
\begin{center}
\epsfxsize=8cm
\begin{minipage}{\epsfxsize}\epsffile{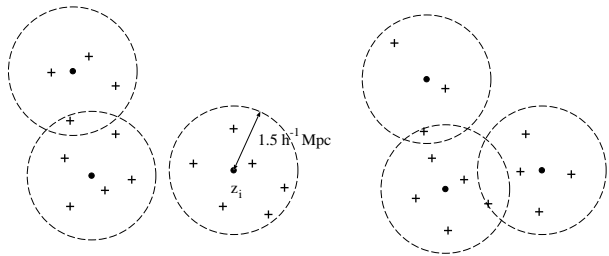}\end{minipage}
\end{center}
\caption{
\label{fig:matern}
Sketch  of a two--dimensional  Mat\'ern  cluster process  with cluster
radius $R=1.5\hMpc$ and mean number of clusters $\mu=5$.}
\end{figure}

Before attempting to  apply $J(r)$ to galaxy  samples, we want to test
it  on a model with  non--trivial yet analytically tractable behaviour
of $J(r)$.

In   order    to   describe     the      clustering    of    galaxies,
{}\scite{neyman:statistical} suggested a class of point processes that
was   subsequently named after   them.  We concentrate   on a subclass
called Mat\'ern cluster   processes.   They are constructed   by first
distributing   uniformly  $M$ cluster  centres.    Around each cluster
centre, which is itself not  included in the final point distribution,
$m$ galaxies are placed randomly within a  sphere of radius $R$, where
$m$ is a   Poisson distributed random variable with   mean  $\mu$.  In
Figure~\ref{fig:matern} we show a sketch of such a process.  Note that
overlapping clusters are allowed.

For  a Mat\'ern cluster  process, {}\scite{vanlieshout:j}  proved that
$J(r)$ is monotonically  decreasing from  1   at $r=0$ and attains   a
constant value for $r>2R$, where $R$ is the radius of a cluster.  This
constant  value  can  be   interpreted  as a   relic   of  the uniform
distribution of the cluster centres.
In three dimensions
\begin{equation} \label{eq:j_matern_1}
J_\M(r) = \left\{
\begin{array}{ll}
\frac{1}{\Vol(B_R)} \int_{B_R}\e^{-\mu V(\bx,r,R)}\d^3x
& \mbox{for}~0{\leq}r{\leq}2R, \\
\exp(-\mu)  &\mbox{for}~r>2R,
\end{array} \right.
\end{equation}
where
\begin{equation}
V(\bx,r,R) = \frac{\Vol(B_r(\bx)\cap B_R)}{\Vol(B_R)}
\end{equation}
denotes the  ratio of the volume  of the intersection of  two balls to
the volume of a single ball.   Here $B_r(\bx)$ is a ball of radius $r$
centred  at the  point $\bx$,  while  $B_R$ is  a ball  of radius  $R$
centred at  the origin.   This quantity can  be calculated  from basic
geometric considerations, both  in two {}\cite{stoyan:fractals} and in
three dimensions, where the result is
\begin{equation}
V(x,r,R) = \left\{
\begin{array}{ll}
c_3x^3+c_1x+c_0+c_{-1}x^{-1} 
& \mbox{for}~0{\leq}r{<}R~\mbox{and}~R-r{<}x{<}R, \\
& \mbox{or}~R{\leq}r{\leq}2R~\mbox{and}~r-R{<}x{<}r, \\
r^{3}/R^{3} 
& \mbox{for}~0{\leq}r{<}R~\mbox{and}~0{\leq}x{\leq}R{-}r, \\
1
& \mbox{for}~R{\leq}r{\leq}2R~\mbox{and}~0{\leq}x{\leq}r{-}R.
\end{array} \right.
\end{equation}
with $x=|\bx|$ and
\begin{equation}
\begin{array}{llll}
c_{3}  = \frac{1}{16 R^{3}}, &
c_{1}  = -\frac{3}{8} \big(\frac{r^{2}}{R^{3}} + \frac{1}{R}\big), &
c_{0}  = \frac{1}{2}  \big(\frac{r^{3}}{R^{3}} + 1\big), &
c_{-1} = \frac{3}{16} \big(\frac{2 r^2}{R} - \frac{r^4}{R^3} -  R\big) ,
\end{array}
\end{equation}

\begin{figure}
\begin{center}
\epsfxsize=8cm
\begin{minipage}{\epsfxsize}\epsffile{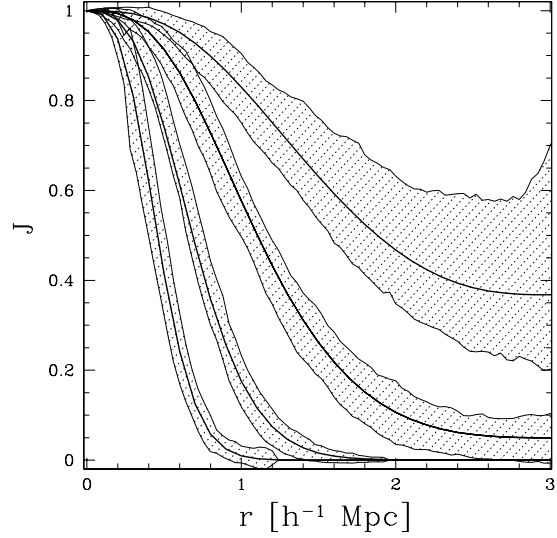}\end{minipage}
\end{center}
\caption{
\label{fig:Jmatern} 
In this panel we show $J_\M$ for Mat\'ern cluster processes with fixed
cluster radius $R=1.5\hMpc$  and varying mean  number  of galaxies per
cluster  $\mu=1,3,10,30$  (bending   down successively).   The   areas
indicate 1--$\sigma$ fluctuations of 50 realizations.}
\end{figure}
In  Figure~\ref{fig:Jmatern} we show\footnote{Throughout this article,
distances  are  given in \hMpc,  where $h$  denotes  the value  of the
Hubble  parameter   measured   in   units of  $100\frac{\kms}{\Mpc}$.}
$J_\M(r)$ for  $R=1.5\hMpc$   and  several   values  of   $\mu$;  this
represents typical situations  of galaxy clustering.  Obviously $J(r)$
discriminates between the  varying richness  classes of the   Mat\'ern
cluster processes.

\section{Galaxy samples}
\label{sect:galaxies}

In this section  we want to go one  step further by applying $J(r)$ to
catalogues of  galaxies and groups of  galaxies, and compare them with
a Mat\'ern cluster process.

\subsection{Description of the PPS galaxy and group samples}
\label{sect:pps}

The     PPS   database   was      compiled   in   the   last    decade
({}\pcite{giovanelli:redshift}, {}\pcite{wegner:survey}).  The    full
redshift survey is magnitude--limited down   to a Zwicky magnitude  of
$m_\Z\leq15.7$ {}\cite{zwicky:catalogue},  and at least  95\% complete
to  $m_\Z\leq15.5$   (see  Figure~1 in  {}\pcite{iovino:galaxy}).   We
extract a  volume--limited  subsample  with  $M_\Z\leq-19$ and  radius
$79\hMpc$,   confined  to        $-1^h.50\le\alpha\le+3^h.00$      and
$0\deg\le\delta\le40\deg$, i.e.\ a solid angle of $0.76{\rm sr}$.
Redshifts are corrected for the motion of the Sun relative to the rest
frame    of   the    Cosmic  Microwave   Background     (CMB)   as  in
{}\scite{peebles:principles},  and  we also correct  Zwicky magnitudes
for interstellar extinction as in {}\scite{burstein:HI}.
The final volume--limited sample PPS79 contains 817 galaxies.

To find  groups,  we   use the  redshift   space  friends--of--friends
algorithm of {}\scite{huchra:groups},  suitably adap\-ted to our case.
It  is a truncated percolation algorithm  with two independent linking
parameters  $D$ and  $V$.   Briefly, two  galaxies are  ``friends'' if
their transverse    and   radial    separations $r_{ij}^{\perp}$   and
$r_{ij}^{\parallel}$   satisfy       $r_{ij}^{\perp}     \leq  D$  and
$r_{ij}^{\parallel}     \leq V/H_0$,   respectively.  Friendship    is
transitive, and a set of {\em three} or more friends is called a loose
group of galaxies.

Usually, loose   groups are identified in  magnitude--limited samples.
Here,  we  consider    only   volume--limited   samples.    Values  of
$D=0.52\hMpc$  and $V=600~\kms$  give  very good  agreement of  global
properties (e.g.\ the total fraction  of galaxies in groups, the ratio
of groups to galaxies, or the  median velocity dispersion) between our
volume--limited group catalogue  and the  magnitude--limited catalogue
constructed by {}\scite{trasarti:loose}.

The final sample contains 230 galaxies  in 48 loose groups.  A typical
group    has 5  observed   members,    a  ``virial   mass''   of  some
$10^{13}M_\odot$, and an observed luminosity of some $10^{10}L_\odot$.
Both its  radius and its  inter--member pairwise separation are around
$0.5\hMpc$, and  the  line--of--sight velocity  dispersion amounts  to
roughly $200~\kms$,   so  the groups   appear  thin  and elongated  in
redshift space.

\subsection{$J(r)$ for the  galaxy samples}
\label{sect:pps_groups}

\begin{figure}
\begin{center}
\epsfxsize=8cm
\begin{minipage}{\epsfxsize}\epsffile{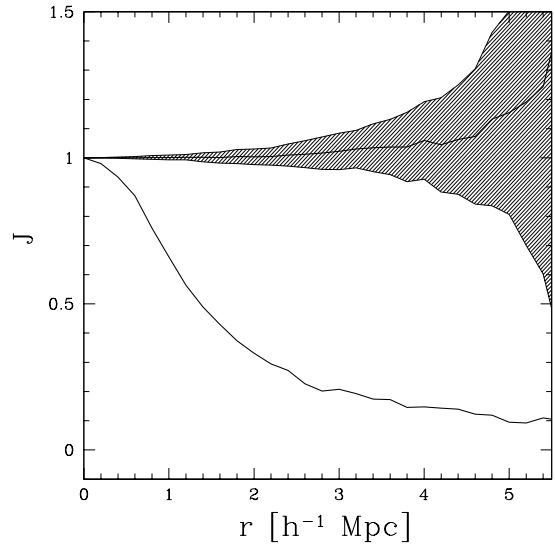}\end{minipage}
\end{center}
\caption{
\label{fig:J_PPS_GAL} 
$J(r)$ for  all galaxies in the  PPS79 sample (solid line)  and for 50
realizations  of a  Poisson process  with  the same  number of  points
(dashed area).  The dashed area indicates 1--$\sigma$ fluctuations.}
\end{figure}
We calculated $J(r)$   for all galaxies from   the PPS79  sample;  the
results  are  shown in Figure~\ref{fig:J_PPS_GAL}.   With $J(r)$ lying
outside the area  occupied by realizations  of a Poisson process,  one
can  clearly see   that galaxies  are   strongly clustered  -- not   a
particularly surprising result.    In Sect.~\ref{sect:PPS-simulation},
somewhat  more   interesting  comparisons   with  galaxy mock--samples
extracted from $N$--body simulations are performed.

\begin{figure}
\begin{center}
\epsfxsize=8cm
\begin{minipage}{\epsfxsize}\epsffile{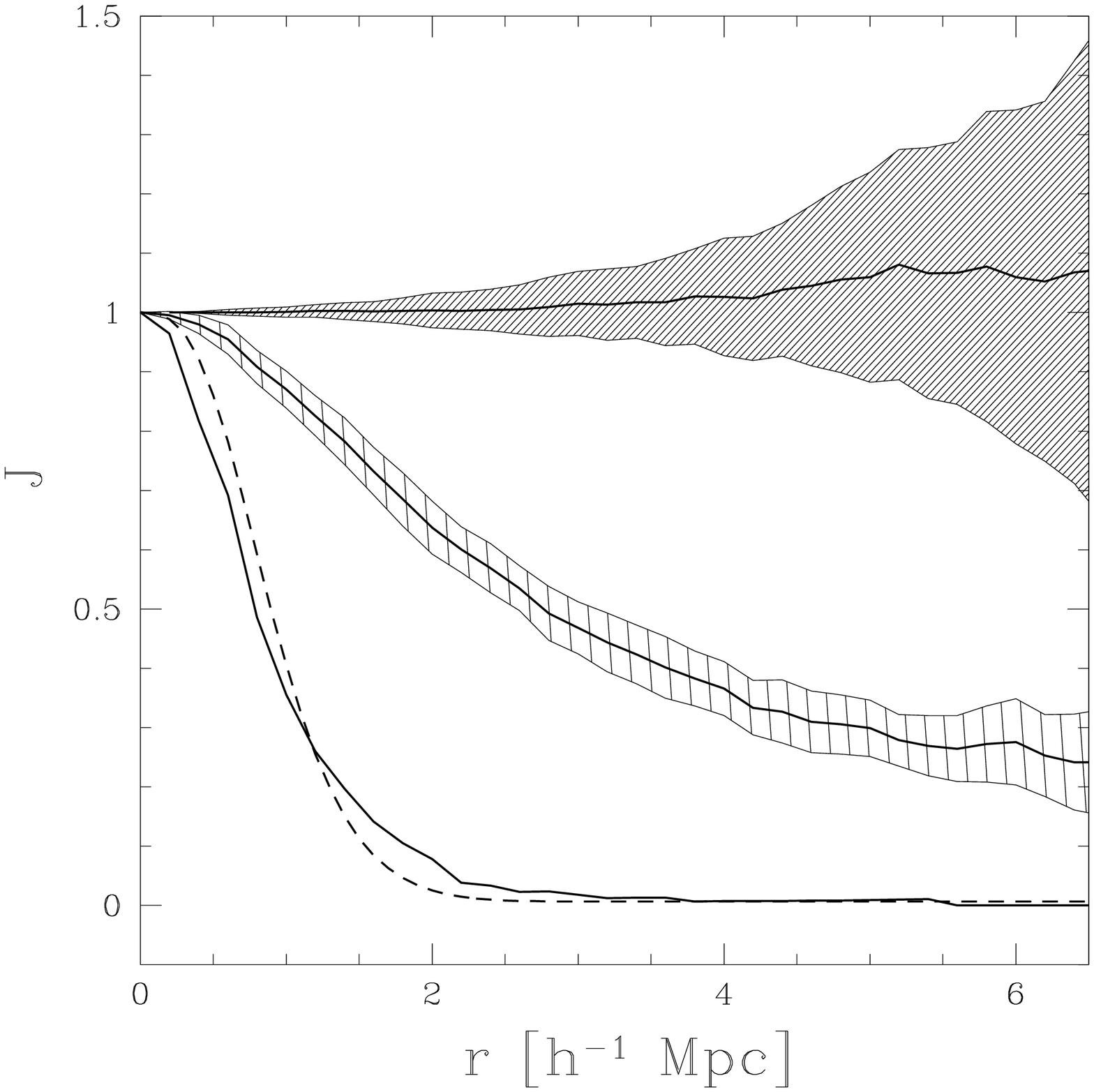}\end{minipage}
\end{center}
\caption{
\label{fig:J_PPS_GR} 
$J(r)$  for the galaxies in  groups in the  PPS79 sample (solid line),
for a Mat\'ern  cluster process with  $\mu=5$ and $R=1.5\hMpc$ (dashed
line), for the average of 50 samples extracted  from all galaxies with
the same number density as the galaxies in groups (light shaded area),
and for the average  and fluctuations of  50 realizations of a Poisson
process (dark shaded area).}
\end{figure}
Figure~\ref{fig:J_PPS_GR}      displays  the   results   for   grouped
galaxies. Since  each   group contains  at  least  three  members, the
nearest neighbour  of a grouped  galaxy is  certainly found within the
largest link length used in the friends--of--friends procedure.  Hence
we observe $G(r)=1$ and subsequently  $J(r)=0$ for $r>5.6\hMpc$ in the
grouped galaxy sample.
$J(r)$ is in general {\em  not} invariant under  changes of the number
density {}\cite{vanlieshout:j}.   To compare  the $J(r)$  for  grouped
galaxies with the $J(r)$   for all galaxies,  we subsample  the denser
PPS79.  $J(r)$   is  calculated from  50   subsamples of 230  galaxies
randomly selected from the whole PPS79  sample. With $J(r)$ we measure
the strength   of  clustering, which  is  emphasised when  we consider
galaxies  in groups only,  and is less  pronounced when we look at the
whole sample with field galaxies included.
Similarly the  value of  $J(r)$ for the  sub--sampled PPS79  is higher
than  $J(r)$  for  the   whole  PPS79,  because  Random  sub--sampling
(thinning) tends to increase $J(r)$ towards the Poisson value.

The  centers  of loose  groups  show  a strong  correlation themselves
{}\cite{trasarti:clustering}, therefore a Mat\'ern cluster process can
only serve   as  a rough approximation   to  the true  distribution of
galaxies  in groups.  Despite  this,  a Mat\'ern cluster process  with
$\mu=5$  galaxies   per  group   (cluster)  and  a  group    radius of
$R=1.5\hMpc$ shows a $J(r)$ comparable to the $J(r)$ obtained from the
galaxies in groups, where  in the mean  4.8 galaxies reside in a group
(see Fig.~\ref{fig:Jmatern}).
We see a low, almost constant value of $J(r)$  for $r>2.5\hMpc$.  This
suggests that we are indeed looking  at highly clustered galaxies with
small contamination by ``field'' galaxies.

The $J_M(r)$  of a Mat\'ern  cluster process  gets constant  for radii
twice      as     large   as   the       cluster   radius.    Already,
{}\scite{vanlieshout:j} express their hope  to deduce a cluster  scale
$R$ in a point distribution  from $J(2R)\approx{\rm const}$.  However,
this must be    taken with extreme caution.     As can  be seen   from
Fig.~\ref{fig:Jmatern} we may be  fooled by a  factor of three by  the
fluctuations in the estimated $J(r)$.    The uncertainty becomes  even
worse when we  consider certain Cox--processes, where $J(r)$ decreases
strictly     monotonically     towards      a       constant     value
{}\cite{vanlieshout:j}, and in principle  no scale can be deduced from
the  comparison  with the   oversimplified  Mat\'ern cluster  process.
Either we have to   restrict ourselves to qualitative  statements,  or
come up with more refined and realistic models.

\section{Comparison with  N--body simulations}
\label{sect:nbody}

The  preceding section  showed that  the qualitative  features  of the
galaxy distribution are well  described by the $J$--function.  In this
section we  explicate that  the $J$--function is  also suitable  for a
quantitative  comparison,  and  allows  us to  constrain  cosmological
models.

\subsection{N--body simulations and mock--catalogues}
\label{sect:simulation}

We extract 64    mock--PPS  catalogues from a    cosmological  N--body
simulation of a Mixed Dark Matter (MDM) model.

We consider  a   MDM model with  one   species  of  massive neutrinos,
dimensionless  Hubble   parameter $h  =  0.5$  and  density parameters
$\Omega_c = 0.8$,  $\Omega_h  = 0.2$ for  cold  and hot  dark  matter,
respectively.  The analytical   expressions for the MDM  power spectra
$P(k)$  was taken from  {}\scite{ma:linear}.   The initial $P(k)$  was
normalised to the   COBE 4-yr  data {}\cite{bunn:fouryear}, giving   a
corresponding   value  of $\sigma_8 =  0.82$    for the  r.m.s.\  mass
fluctuation in an $8\hMpc$ sphere.

The simulation was  run from an initial  expansion factor $a_i=1$ down
to $a_f=4.5$ using a P$^3$M code  with $100^3$ particles of mass $1.49
\cdot 10^{13} M_\odot$, on a cubic grid of $256^3$ cells, with a force
softening radius  $0.32\hMpc$,  in  a box   of  side $300\hMpc$.   The
integration was performed in comoving coordinates using $a(t)$ as time
variable for a total of 225 steps.

We identify ``galaxies'' in our   simulation with a method similar  to
the one discussed by {}\scite{little:cosmic}:

First, we associate with each particle a number $n_i$ of galaxy--scale
peaks   calculated     from  the   initial   density   contrast  field
$\delta(\bx)$.    In    the   peak--background  split    approximation
(\pcite{bardeen:gauss},                      {}\pcite{white:clusters},
{}\pcite{park:biasing})   $n_i$ is the  number   of galaxy peaks  with
height $\delta_s(\bx_i) \geq \nu_{th} \sigma_s$, where $\delta_s(\bx)$
denotes  the  field    smoothed   with a   Gaussian    kernel of width
$R_s=0.55\hMpc$, and $\sigma_s^2$ gives the smoothed field's variance.
The  field is  subject   to the constraint  that   it takes the  value
$\nu_b\sigma_b$      when   smoothed  on  a       scale $R_b>R_s$ (see
{}\pcite{park:biasing}   for more details).  Choosing $\nu_{th}=0.05$,
at $a=4.5$  the  particle  two--point correlation  function,  weighted
according   to $n_i$, matches     in slope and  amplitude the   galaxy
two--point correlation function.    For  the adopted   parameters, the
total number of peaks in the box is $\sum n_i \simeq 690,000$.

Then,  we select the  $i$--th particle as a  galaxy if $An_i>p$, where
$p\in (0,1)$ is a uniformly distributed  random variable, and $A$ is a
constant of proportionality.   The  latter is  set by the  requirement
that the mean  number density of ``galaxies'' in  the box matches  the
mean   density of $M\leq  -19+5   \log(h)$ galaxies  expected from the
Schechter  luminosity function   with  $\alpha=-1.15$, $M_*=-19.3  + 5
\log(h)$,  $\phi_*=0.02  \,  h^{3}$Mpc$^{-3}$   appropriate   for  PPS
(\pcite{trasarti:loose};       {}\pcite{marzke:luminosity}).      This
Monte--Carlo procedure makes the  implicit assumption that the  higher
the peak, the more luminous the associated galaxy.

The mock--PPS  catalogues were built as follows.   The simulation cube
was  divided into $64$  sub--cubes of  side length  $75\hMpc$.  Within
each  sub--cube  we  fit   a  PPS--like  wedge  of  radius  $79\hMpc$.
Redshift--space coordinates $\alpha$,  $\delta$, $cz$ were assigned to
all the  ``galaxies'' of  the sub--cubes.  Finally,  we kept  only the
``galaxies'' within  the redshift--space--boundaries of  the mock--PPS
catalogues.

\begin{figure}
\begin{center}
\epsfxsize=8cm
\begin{minipage}{\epsfxsize}\epsffile{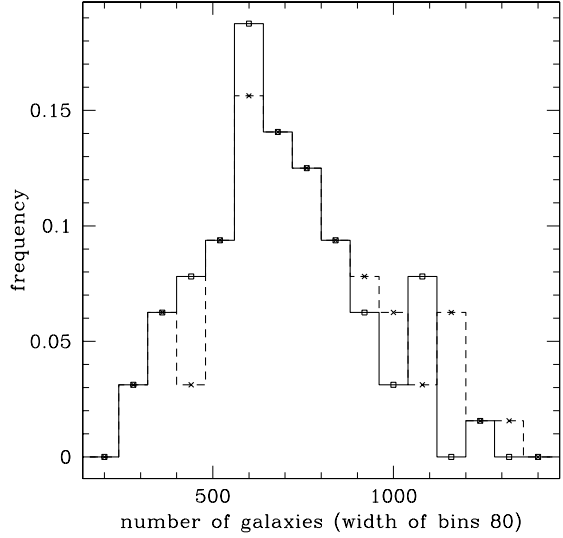}\end{minipage}
\end{center}
\caption{
\label{fig:counts_histogram} 
This   histogram displays the number  of  points per mock--sample; the
solid lines give the values in redshift--space, while the dashed lines
corresponds to selection in real--space.}
\end{figure}

Although we are looking at a large volume with a  depth 79\hMpc\ and a
solid  angle of $0.76{\rm sr}$,  we observe large fluctuations of 25\%
in         the     number     of      points     per      mock--sample
(Fig.~\ref{fig:counts_histogram}).     This  is  consistent  with  the
large--scale   fluctuations  of  the  clustering   properties  of IRAS
galaxies, as found by {}\scite{kerscher:fluctuations} out to scales of
200\hMpc, and  expresses cosmic  variance  in agreement  with expected
sample--to--sample variations  {}\cite{buchert:two-point}.  As we will
see, this slightly complicates the analysis.

\subsection{$J(r)$ for the mock--samples}
\label{sect:J-mock}

\begin{figure}
\begin{center}
\epsfxsize=8cm
\begin{minipage}{\epsfxsize}\epsffile{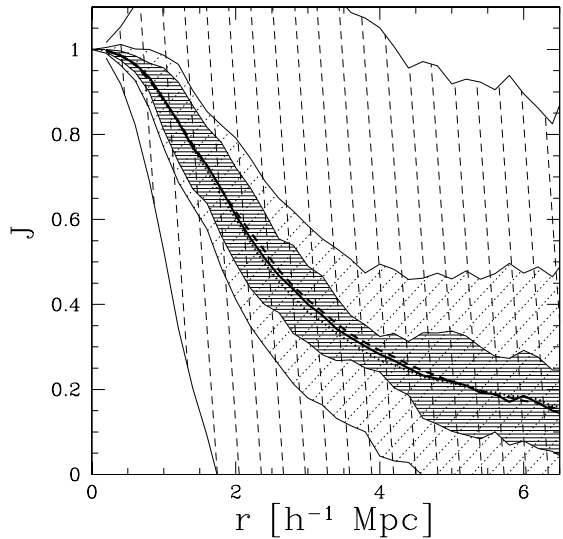}\end{minipage}
\end{center}
\caption{
\label{fig:red-all-simul} 
The mean $J(r)$ and  the 1$\sigma$  range  for the mock--samples  with
$\Delta=30$ (solid line, dark shaded), with $\Delta=100$ (dotted line,
medium shaded), and for  all mock--samples (dashed line, light shaded,
the 1$\sigma$ range is plotted symmetrically).}
\end{figure}
At first we investigate the mock--samples selected in redshift--space.
If  we  use  all  the   64  mock--samples  we  are  dominated  by  the
fluctuations  between samples  with  a different  number density  (see
Fig.~\ref{fig:red-all-simul}).   Therefore, we  restrict  ourselves to
mock--samples with approximately  the same number of points  as in the
observed  galaxy  sample: $N_{\rm  gal}-\Delta  \le  N_{\rm mock}  \le
N_{\rm gal}+\Delta$, with $N_{\rm gal}=817$.  For $\Delta=30$ only six
samples  enter, whereas  for  $\Delta=100$ we  already have  seventeen
mock--samples  to analyse.  The  mean value  of $J(r)$  hardly changes
between samples with different  $\Delta$.  Obviously, samples with low
density  tend  to  be  centred  on voids,  and  high--density  samples
typically include large, Coma--like clusters. So large fluctuations in
the  number  density lead  to  large  fluctuations  in the  clustering
properties  measured  by  $J(r)$   but  cancel  in  the  mean.   These
fluctuations      decrease     for      smaller      $\Delta$     (see
Fig.~\ref{fig:red-all-simul}; this was confirmed by inspecting samples
with $\Delta=50$ and $\Delta=200$).
In  order  to look  at  structures comparable  to  the  PPS sample  we
consider  mock--samples  with  a  similar  number density  as  in  the
observed galaxy  sample, and do  not subsample the  mock--samples with
high number density.

\begin{figure}
\begin{center}
\epsfxsize=8cm
\begin{minipage}{\epsfxsize}\epsffile{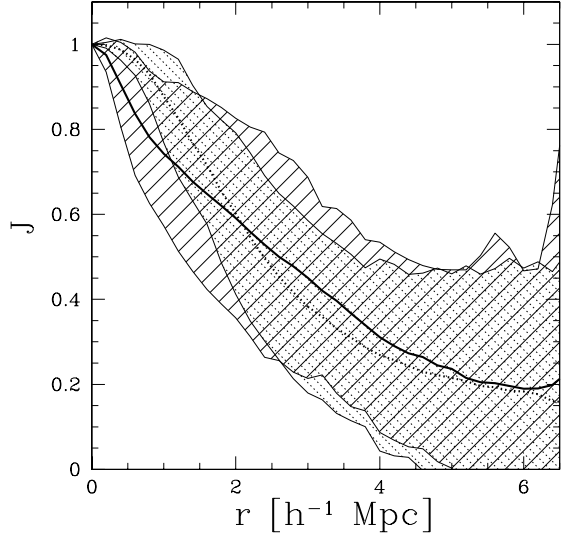}\end{minipage}
\end{center}
\caption{
\label{fig:real-red-data} 
This plot shows  the average value of $J(r)$  and  1$\sigma$ range for
the mock--samples with $\Delta=100$  in real--space (solid lines), and
for the mock--samples in redshift--space (dotted lines).}
\end{figure}
In Fig.~\ref{fig:real-red-data}  the results  of the  mock--samples in
real-- and  redshift--space are compared.  The  mock--samples selected
in   redshift--space  show a   weaker  clustering  than  mock--samples
selected in real--space on small  scales out to  at least $2\hMpc$, as
can be deduced from  the higher  $J(r)$.   This can be traced  back to
redshift space distortions.
The peculiar motions act  by erasing small scale clustering; therefore
the J value of redshift-space samples is larger (less clustering) than
that of  real-space samples. This  effect changes at a  given distance
(2\hMpc). The same  effect  was found by {}\scite{martinez:galaxy}  in
volume--limited subsamples,   extracted  from CfA-I, by  means  of the
two--point correlation function.

\subsection{Comparison of the PPS galaxies with the mock--samples}
\label{sect:PPS-simulation}

\begin{figure}
\begin{center}
\epsfxsize=8cm
\begin{minipage}{\epsfxsize}\epsffile{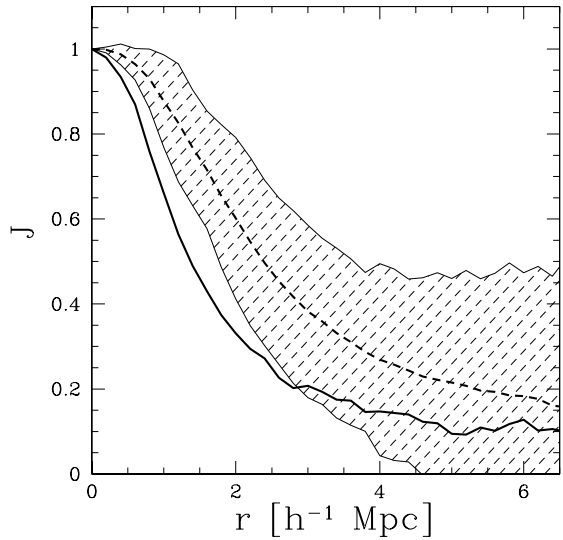}\end{minipage}
\end{center}
\caption{
\label{fig:redshift-data} 
$J(r)$ is   shown for the PPS79   galaxy sample (solid   line) and the
1$\sigma$ range   for   the   mock--samples   with   $\Delta=100$   in
redshift--space (dashed area).}
\end{figure}
In Fig.~\ref{fig:redshift-data}   the results of  the mock--samples in
redshift--space  are compared with the  results of the observed galaxy
distribution  in  the   PPS.    The mock--samples show    insufficient
clustering on small scales out to at least $3\hMpc$, as can be deduced
from the higher $J(r)$.   This is  probably  due to the high  velocity
dispersion in  MDM models  {}\cite{jing:largescale}.  In  real--space,
which  is {\em  not}  directly  comparable   with  the  PPS data,  the
mock--samples reproduce the clustering on small  scales out to 1\hMpc,
but  again show   not  enough  clustering,   even though they   become
marginally consistent with the  observed galaxy distribution on larger
scales.
We have  to conclude that this  MDM simulation is  unable to reproduce
the observed strong clustering of galaxies on small scales.  Of course
this result depends on our method of galaxy identification. A different
biasing prescription might change this. 
On large  scales a  definitive answer is  not possible, since  for $r$
larger than  6\hMpc\ an estimation  of $J(r)$ becomes  unreliable; the
empirical $G(r)$ and $F(r)$ approach unity, and the quotient $J(r)$ is
ill--defined.

\section{Conclusion and Outlook}
\label{sect:outlook}

We  have highlighted promising properties  of the global morphological
descriptor $J(r)$.  It  connects the distribution functions $F(r)$ and
$G(r)$ and, hence,  incorporates all orders of correlation  functions.
$J(r)$ measures   the strength of  clustering  in a point  process and
distinguishes between correlated  and  anti--correlated patterns.  
The  example of  a  Mat\'ern cluster  process  illustrates that $J(r)$
sensitively depends on the richness of the clusters or groups.

Since $J(r)$ is built  from  cumulated distribution functions, we  do
not   encounter spurious  results   due   to  binning.  This   becomes
particularly important on small scales.

The application  of the $J$-function  to galaxies in a  volume limited
sample and to a sample of  galaxies in loose groups clearly showed the
stronger clustering  of galaxies  in groups.  In  a comparison  with a
Mat\'ern cluster we found  that internal properties, like the richness
of  loose  groups,  are  satisfactorily modelled.   However,  for  the
large--scale  distribution of galaxies,  the Mat\'ern  cluster process
clearly is an over--simplification.

We used   the $J$-function for  a  comparison  of the  observed galaxy
distribution with  galaxy  mock--samples.  Although  the mock--samples
extracted from   a MDM--simulation cover  a large  volume, we detected
large  fluctuations of the  order of 25\%  in the number of points per
sample.  On small scales, out to 1\hMpc, the clustering in real--space
is   as strong as    in the  observed galaxy   distribution,   but the
comparable redshift--space mock--samples show too weak clustering.  On
larger  scales   from  2--6\hMpc\   both real--   and  redshift--space
mock--samples show too weak clustering.  Hence, this MDM simulation is
not able  to reproduce the  observed strong clustering of the galaxies
on small scales.

The function $J(r)$  has  proved to achieve comparable  discriminative
power as the Minkowski functionals {}\cite{kerscher:fluctuations}, and
is  most suitable for  addressing  the  question of ``regularity''  on
large--scales  as demonstrated in  an analysis of  the distribution of
superclusters {}\cite{kerscher:regular}.
In  this article we have  shown that the   $J(r)$ function is a useful
tool for quantifying the clustering of galaxies on small scales and is
capable of constraining cosmological models of structure formation.

\section*{Acknowledgements}

It is a pleasure to thank Adrian Baddeley, Bryan Scott, Claus Beisbart
and  Herbert Wagner for   useful discussions  and comments.   We thank
Simon D.M.~White  for pointing  out the  relation to   the conditional
correlation function.   This  work was partially   supported by the EC
network of the program Human Capital and Mobility No.\ CHRX-CT93-0129,
the  Acci\'on     Integrada  Hispano--Alemana HA-188A    (MEC),    the
Sonderforschungsbereich SFB     375   f\"ur   Astroteilchenphysik  der
Deutschen Forschungsgemeinschaft, and the Spanish DGES (project number
PB96-0797).


\begin{thebibliography}{\protect\citefmt{Trasarti-Battistoni \bgroup et
  al.\egroup }{1997}}

\bibitem[\protect\citefmt{Bardeen \bgroup et al.\egroup }{1986}]{bardeen:gauss}
Bardeen, J.~M., Bond, J.~R., Kaiser, N., \& Szalay, A.~S.,
  \providecommand{\apj}{Ap.\ J.}{\apj} \textbf{304} (1986), 15

\bibitem[\protect\citefmt{Buchert \& Mart{\'\i}nez}{1993}]{buchert:two-point}
Buchert, T. \& Mart{\'\i}nez, V.~J., \providecommand{\apj}{Ap.\ J.}{\apj}
  \textbf{411} (1993), 485

\bibitem[\protect\citefmt{Bunn \& White}{1997}]{bunn:fouryear}
Bunn, E.~F. \& White, M., \providecommand{\apj}{Ap.\ J.}{\apj} \textbf{480}
  (1997), 6

\bibitem[\protect\citefmt{Burstein \& Heiles}{1978}]{burstein:HI}
Burstein, D. \& Heiles, C., \providecommand{\apj}{Ap.\ J.}{\apj} \textbf{225}
  (1978), 40

\bibitem[\protect\citefmt{Gazta{\~n}aga \& Frieman}{1994}]{gaztanaga:bias}
Gazta{\~n}aga, E. \& Frieman, J.~A., \providecommand{\apjl}{Ap.\ J.\
  Lett.}{\apjl} \textbf{437} (1994), L13

\bibitem[\protect\citefmt{Giovanelli \& Haynes}{1991}]{giovanelli:redshift}
Giovanelli, R. \& Haynes, M.~P., ARA\&A \textbf{29} (1991), 499

\bibitem[\protect\citefmt{Huchra \& Geller}{1982}]{huchra:groups}
Huchra, J. \& Geller, M., \providecommand{\apj}{Ap.\ J.}{\apj} \textbf{257}
  (1982), 423

\bibitem[\protect\citefmt{Iovino \bgroup et al.\egroup }{1993}]{iovino:galaxy}
Iovino, A., Giovanelli, R., Haynes, M.~P., Chincarini, G., \& Guzzo, L.,
  \providecommand{\mnras}{Mon.\ Not.\ Roy.\ Astron.\ Soc.}{\mnras} \textbf{265}
  (1993), 21

\bibitem[\protect\citefmt{Jing \bgroup et al.\egroup }{1994}]{jing:largescale}
Jing, Y., Mo, H., B{\"o}rner, G., \& Fang, L.~Z.,
  \providecommand{\aanda}{Astron.\ Astrophys.}{\aanda} \textbf{284} (1994),
  703

\bibitem[\protect\citefmt{Kerscher \bgroup et al.\egroup
  }{1997}]{kerscher:abell}
Kerscher, M., Schmalzing, J., Retzlaff, J., Borgani, S., Buchert, T.,
  Gottl{\"o}ber, S., M{\"u}ller, V., Plionis, M., \& Wagner, H.,
  \providecommand{\mnras}{Mon.\ Not.\ Roy.\ Astron.\ Soc.}{\mnras} \textbf{284}
  (1997), 73

\bibitem[\protect\citefmt{Kerscher \bgroup et al.\egroup
  }{1998}]{kerscher:fluctuations}
Kerscher, M., Schmalzing, J., Buchert, T., \& Wagner, H.,
  \providecommand{\aanda}{Astron.\ Astrophys.}{\aanda} \textbf{333} (1998),
  1

\bibitem[\protect\citefmt{Kerscher}{1998}]{kerscher:regular}
Kerscher, M., \providecommand{\aanda}{Astron.\ Astrophys.}{\aanda} \textbf{336}
  (1998), 29

\bibitem[\protect\citefmt{Little \& Weinberg}{1994}]{little:cosmic}
Little, B. \& Weinberg, D., \providecommand{\mnras}{Mon.\ Not.\ Roy.\ Astron.\
  Soc.}{\mnras} \textbf{267} (1994), 605

\bibitem[\protect\citefmt{Ma}{1996}]{ma:linear}
Ma, C.-P., \providecommand{\apj}{Ap.\ J.}{\apj} \textbf{471} (1996), 13

\bibitem[\protect\citefmt{{Mart{\'\i}nez} \bgroup et al.\egroup
  }{1990}]{martinez:clustering}
{Mart{\'\i}nez}, V.~J., Jones, B.~J.~T., Dominguez-Tenreiro, R., \& {van de
  Weygaert}, R., \providecommand{\apj}{Ap.\ J.}{\apj} \textbf{357} (1990),
  50

\bibitem[\protect\citefmt{{Mart{\'\i}nez} \bgroup et al.\egroup
  }{1993}]{martinez:galaxy}
{Mart{\'\i}nez}, V.~J., Portilla, M., Jones, B.~J.~T., \& Paredes, S.,
  \providecommand{\aanda}{Astron.\ Astrophys.}{\aanda} \textbf{280} (1993),
  5

\bibitem[\protect\citefmt{Marzke \bgroup et al.\egroup
  }{1994}]{marzke:luminosity}
Marzke, R., Huchra, J., \& Geller, M., \providecommand{\apj}{Ap.\ J.}{\apj}
  \textbf{428} (1994), 43

\bibitem[\protect\citefmt{Mecke \bgroup et al.\egroup }{1994}]{mecke:robust}
Mecke, K.~R., Buchert, T., \& Wagner, H., \providecommand{\aanda}{Astron.\
  Astrophys.}{\aanda} \textbf{288} (1994), 697

\bibitem[\protect\citefmt{Melott}{1990}]{melott:review}
Melott, A.~L., Physics Rep. \textbf{193} (1990), 1

\bibitem[\protect\citefmt{Neyman \& Scott}{1958}]{neyman:statistical}
Neyman, J. \& Scott, E.~L., J.\ R.\ Stat.\ Soc. \textbf{20} (1958), 1

\bibitem[\protect\citefmt{Park}{1991}]{park:biasing}
Park, C., \providecommand{\mnras}{Mon.\ Not.\ Roy.\ Astron.\ Soc.}{\mnras}
  \textbf{251} (1991), 167

\bibitem[\protect\citefmt{Peebles}{1993}]{peebles:principles}
Peebles, P.~J.~E., \emph{Principles of physical cosmology}, Princeton
  University Press, Princeton, New Jersey, 1993

\bibitem[\protect\citefmt{Schmalzing \& Buchert}{1997}]{schmalzing:beyond}
Schmalzing, J. \& Buchert, T., \providecommand{\apjl}{Ap.\ J.\ Lett.}{\apjl}
  \textbf{482} (1997), L1

\bibitem[\protect\citefmt{Sharp}{1981}]{sharp:holes}
Sharp, N., \providecommand{\mnras}{Mon.\ Not.\ Roy.\ Astron.\ Soc.}{\mnras}
  \textbf{195} (1981), 857

\bibitem[\protect\citefmt{Stoyan \& Stoyan}{1994}]{stoyan:fractals}
Stoyan, D. \& Stoyan, H., \emph{Fractals, Random Shapes and Point Fields}, John
  Wiley \& Sons, Chichester, 1994

\bibitem[\protect\citefmt{Stoyan \bgroup et al.\egroup
  }{1995}]{stoyan:stochgeom}
Stoyan, D., Kendall, W.~S., \& Mecke, J., \emph{Stochastic Geometry and its
  Applications}, 2nd ed., John Wiley \& Sons, Chichester, 1995

\bibitem[\protect\citefmt{Stratonovich}{1963}]{stratonovich:topicsI}
Stratonovich, R.~L., \emph{Topics in the theory of random noise}, Vol.~1,
  Gordon and Breach, New York, 1963

\bibitem[\protect\citefmt{Trasarti-Battistoni \bgroup et al.\egroup
  }{1997}]{trasarti:clustering}
Trasarti-Battistoni, R., Invernizzi, G., \& Bonometto, S.~A.,
  \providecommand{\apj}{Ap.\ J.}{\apj} \textbf{475} (1997), 1

\bibitem[\protect\citefmt{Trasarti-Battistoni}{1998}]{trasarti:loose}
Trasarti-Battistoni, R., \providecommand{\aas}{Astron.\ Astrophys.\
  Suppl.}{\aas} \textbf{130} (1998), 341

\bibitem[\protect\citefmt{{van Lieshout} \& Baddeley}{1996}]{vanlieshout:j}
{van Lieshout}, M.~N.~M. \& Baddeley, A.~J., Statist.\ Neerlandica \textbf{50}
  (1996), 344

\bibitem[\protect\citefmt{Wegner \bgroup et al.\egroup }{1993}]{wegner:survey}
Wegner, G., Haynes, M.~P., \& Giovanelli, R., \providecommand{\aj}{A.\ J.}{\aj}
  \textbf{105} (1993), 1251

\bibitem[\protect\citefmt{White \bgroup et al.\egroup }{1987}]{white:clusters}
White, S.~D.~M., Frenk, C.~S., Davis, M., \& Efstathiou, G.,
  \providecommand{\apj}{Ap.\ J.}{\apj} \textbf{313} (1987), 505

\bibitem[\protect\citefmt{White}{1979}]{white:hierarchy}
White, S.~D.~M., \providecommand{\mnras}{Mon.\ Not.\ Roy.\ Astron.\
  Soc.}{\mnras} \textbf{186} (1979), 145

\bibitem[\protect\citefmt{Zwicky \bgroup et al.\egroup
  }{1968}]{zwicky:catalogue}
Zwicky, F., Herzog, E., Wild, P., Karpowicz, M., \& Kowal, C., \emph{Catalogue
  of Galaxies and Clusters of Galaxies}, Vol. 1--6, California Inst. of
  Technology, Pasadena, 1961--1968

\end{thebibliography}

\providecommand{\bysame}{\leavevmode\hbox to3em{\hrulefill}\thinspace}

\end{document}